\documentclass{sig-alternate}
\usepackage{verbatim, times,amsmath,epsfig} 
\usepackage{scrextend} 
\usepackage[vlined,linesnumbered, ruled, commentsnumbered]{algorithm2e}
\usepackage{float}
\usepackage{multicol}
\usepackage{flushend}
\usepackage{cite}
\usepackage{color}
\usepackage{graphicx}
\usepackage{mathrsfs}
\usepackage{relsize}
\usepackage{url}
\usepackage[bottom]{footmisc}
\usepackage{caption}
\DeclareCaptionType{copyrightbox}
\usepackage{subcaption}
\usepackage{bm}

\newcommand{\balpha}{\bm{\alpha}}
\newcommand{\bw}{{\mathbf w}}
\newcommand{\bx}{{\mathbf x}}
\newcommand{\by}{{\mathbf y}}
\newcommand{\bz}{{\mathbf z}}
\newcommand{\ba}{{\mathbf a}}
\newcommand{\bb}{{\mathbf b}}
\newcommand{\Lag}{\mathscr{L}}

\title{Fast Support Vector Machines Using Parallel Adaptive Shrinking on
Distributed Systems}
\numberofauthors{4}
\author{
\alignauthor 
    Jeyanthi Narasimhan\\
    \affaddr{School of Electrical Engineering and Computer Science,}\\
    \affaddr{Washington State University,}\\
   \affaddr{Pullman, WA 99164}\\
       \texttt{jsalemna@eecs.wsu.edu}
\alignauthor 
    Abhinav Vishnu\\
    \affaddr{Computational Science and Mathematics Division,}\\
    \affaddr{Pacific Northwest National Laboratory,}\\
   \affaddr{ 902 Battelle Blvd, Richland, WA 99352}\\
       \texttt{abhinav.vishnu@pnnl.gov}
\alignauthor 
    Lawrence Holder\\
    \affaddr{School of Electrical Engineering and Computer Science,}\\
    \affaddr{Washington State University,}\\
   \affaddr{Pullman, WA 99164}\\
       \texttt{holder@wsu.edu}
\and
\alignauthor Adolfy Hoisie\\
    \affaddr{Computational Science and Mathematics Division,}\\
    \affaddr{Pacific Northwest National Laboratory,}\\
   \affaddr{ 902 Battelle Blvd, Richland, WA 99352}\\
       \texttt{adolfy.hoisie@pnl.gov}
}
\begin{document}
\toappear{}
\maketitle

\begin{abstract}
Support Vector Machines (SVM), a popular machine learning technique, has
been applied to a wide range of domains such as science, finance, and
social networks for supervised learning.  Whether it is identifying
high-risk patients by health-care professionals, or potential
high-school students to enroll in college by school districts, SVMs can
play a major role for social good.  This paper undertakes the challenge
of designing a scalable parallel SVM training algorithm for large scale
systems, which includes commodity multi-core machines, tightly connected
supercomputers and cloud computing systems. Intuitive techniques for improving
the time-space complexity including adaptive
elimination of samples for faster convergence and sparse format
representation are proposed. Under sample elimination, several heuristics for {\em earliest possible} to
{\em lazy }elimination of non-contributing samples are proposed. In several cases, where an early sample elimination might result
in a false positive, low overhead mechanisms for reconstruction of key
data structures are proposed.  The algorithm and heuristics are
implemented and evaluated on various publicly available datasets. Empirical
evaluation shows up to 26x speed improvement on some datasets against the
sequential baseline, when
evaluated on multiple compute nodes, and an improvement in execution
time up to 30-60\% is readily observed on a number of other datasets against our
parallel baseline.

\end{abstract}

\section{Introduction}
Today, simulations and instruments produce exorbitant amounts of data
and the rate of data production over the years is expected to grow
dramatically~\cite{data:ascac11,data:ascac13}. Machine Learning and Data
Mining (MLDM) provides algorithms and tools for knowledge extraction
from large volumes of data. Several domains such as science, finance and
social networks rely on MLDM algorithms for supervised and unsupervised
learning~\cite{tarca:bio07,vossen:hep08,ml:climate}. Support Vector
Machines (SVM) - a supervised learning algorithm - is ubiquitous due to
excellent accuracy and obliviousness to dimensionality. SVM broadly
relies on the idea of large margin data classification. It
constructs a decision surface in the feature space that bisects the two
categories and maximizes the margin of separation between classes of
points used in the training set. This decision surface is used for
classification on the testing set provided by the user. SVM has strong
theoretical foundations, and the classification and regression
algorithms provide excellent generalization performance~\cite{burges,
nello}. 

With the increasing data volume and general availability of multi-core
machines, several parallel SVM training algorithms are being proposed in
the literature.  PEGASOS~\cite{pegasos} and dual coordinate
descent~\cite{hsieh} train on extremely large problems, albeit with
limitations to
linear SVMs. Cao {\em et al.} have proposed parallel solution extending
the previously proposed Sequential Minimal Optimization (SMO)
algorithm~\cite{keerthi}. However, the empirical evaluation does not
show good scalability and the entire dataset is used for training~\cite{cao:2006:psm}. 
Other algorithms have been proposed for
special architectural aspects such as GPUs~\cite{fsv, cotter:gat}. 
A primary problem with the algorithms proposed above is that they use
the complete dataset for margin generation during the entire
calculation, even though only a fraction of samples (support vectors)
contribute to the hyperplane calculation. 
{\em Shrinking} - a technique to eliminate non-contributing samples - has
been proposed for sequential SVMs~\cite{joachims} to reduce the time
complexity of training. However, no parallel shrinking algorithm for multi-core
machines and distributed systems exists in literature.

This paper addresses the limitations of previously proposed approaches
and provides a novel parallel SVM training algorithm with adaptive
shrinking. We utilize the theoretical framework for shrinking in our parallel
solution to improve on the speed of convergence and use a specific format
for sample representation in the optimization based on the observation that most
of the real world datasets are sparse in nature (see Section~\ref{sec:dso}).
We study the effect of several heuristics (Section~\ref{sec:sh}) for aggressive to conservative
elimination of non-contributing samples during the various stages of execution. The proposed
approaches are designed and implemented uses state-of-the-art
programming models such as Message Passing Interface (MPI)~\cite{mpi1}
and Global Arrays~\cite{nieplocha:psw94} for design of communication and
data storage. These programming models are known to provide optimal
performance on multi-core systems, large scale systems and can be used
on cloud computing systems as well. An empirical evaluation of proposed
approaches shows up to {\bf 3x} speedup in comparison to the original
non-elimination algorithms using the same number of processors, and up
to {\bf 26x}
speedup in comparison to libsvm~\cite{libsvm}. 

\subsection{Contributions}
Specifically, this  paper makes the following contributions:
\begin{enumerate}
		\item 
                Design and analysis of parallel algorithms to improve the
                time complexity of SVM training including adaptive
                elimination of samples. Several heuristics under the categories of
{\em aggressive}, {\em average} and {\em conservative} for elimination of
non-contributing samples.
                \item Space-efficient SVM training algorithm by using compressed
                representation of data samples and avoiding the kernel cache. The
                proposed solution makes it an attractive approach for very large
                -scale datasets and modern systems.
		\item Implementation of our proposed algorithm and evaluation
				with several datasets on multi-core systems and
				large-scale tightly-connected supercomputers. The
				empirical evaluation indicates the efficacy of the
				proposed approach - 5x-8x speedup on USPS and
                                Mushrooms datasets against the sequential
                                baseline~\cite{libsvm} and
                                20-60\% improvement in execution time on several
                                datasets against our parallel no-shrinking
                                baseline algorithm.
\end{enumerate}

The rest of the paper is organized as follows:
section~\ref{sec:background} provides a background of our work.
Section~\ref{sec:design} presents a solution space of the algorithms and
associated heuristics. Empirical evaluation and analysis is performed in
section~\ref{sec:perf}, and section~\ref{sec:related} presents the
related work. Section~\ref{sec:conclusions} presents conclusions and
future directions.

\section{Background}
\label{sec:background}
   Given $\mathcal{N}$ training data points $\{(\bx_1,y_1),
        (\bx_2,y_2), \ldots, (\bx_{\mathcal{N}},y_{\mathcal{N}})\}$
        where $ \bx_i \in {\Re}^d$ and $ y_i \in \{+1, -1\}$, we solve the
        standard two-category soft margin non-linear classification problem.
Thus the problem of finding a maximal margin
separating hyperplane in a high-dimensional space can be formulated as:
\[
\min_{\bw,\beta} \frac{1}{2} \bw\cdot\bw + C(\mathsmaller\sum_i\xi_i)\quad 
\]
\[
\mbox{subject to $y_i ( \bw\cdot\Phi(\bx_i) - \beta) \geq 1-\xi_i$} \quad i=1,\dots,\mathcal{N}
\]
where $C$ is a regularization parameter which is a trade-off between the
classifier generality and its accuracy on the training set, $\xi_i$ is
a positive slack variable allowing noise in the training set and
$\Phi$ maps the input data to a possibly infinite dimensional space
(i.e. $\Phi:\Re^{d}\mapsto\mathcal H$). 

\subsection{SVM Training}
\label{sec:train}
This is a convex quadratic
programming problem~\cite{burges}.  Introducing Lagrange multipliers $\balpha$ and
solving the Lagrangian of the primal to get the Wolfe
dual~\cite{fletcher}, the following formulation is observed:
        \begin{equation}\label{eq:lagdual}
        \max_{\balpha} \Lag_D \equiv \mathsmaller\sum_{i=1}^{\mathcal{N}} \alpha_i - \frac{1}{2}
        \mathsmaller\sum_{i,j=1}^{\mathcal{N}}\alpha_i \alpha_j y_i y_j\Phi(\bx_i) \cdot \Phi(\bx_j)
        \end{equation}
                subject to:
                \begin{equation}\label{eqconst}
        0\leq \alpha_i \leq C, \quad \mathsmaller\sum_i \alpha_i y_i = 0, \quad \forall i=1,\ldots,\mathcal{N}\end{equation}
       Minimizing the primal Lagrangian provides the following formulation: \begin{equation}\label{sol}\bw = \mathsmaller\sum_{i=1}^N \alpha_i y_i \Phi(\bx_i)
       \end{equation}
       
	   SVM training is achieved by a search through the feasible region of the
	   dual problem and maximization of the objective function~\eqref{eq:lagdual},
	   with the Karush-Kuhn-Tucker (KKT) conditions~\cite{burges}
	   in identifying the optimal solution. We refer the reader to~\cite{nello,
    burges} for the full theoretical treatment on the SVMs and training. Samples with $\alpha_i > 0$ are referred as  {\it support vectors}, $\zeta$ (Table~\ref{table:model}). The support vectors  contribute to the
	   definition of the optimal separating hyperplane - other examples
	   can be removed from the dataset. The solution of SV training is
	   given by~\eqref{sol}. 
A new point $\bz$ can be classified with:
\begin{equation}\label{eq:classi}
f(\bz) =  sgn(\bw \cdot \Phi(\bz) - \beta)
          \end{equation}
\subsection{Sequential Minimal Optimization (SMO)}
\label{sec:smo}
SVM training by solving the dual problem is typically conducted by splitting a large optimization
problem into a series of smaller sub-problems~\cite{joachims}. The SMO
algorithm~\cite{plattsmo, keerthi} uses precisely two samples at each
optimization step while solving~\eqref{eq:lagdual}.  This facilitates the generation of an analytical
solution possible for the quadratic minimization at each step because of
the equality constraint in~\eqref{eqconst}. The avoidance of dependencies on
numerical optimization packages makes this algorithm a popular
choice~\cite{libsvm} in SVM training, resulting in simplified design and
reduced susceptibility to numerical issues~\cite{plattsmo}. 
\subsubsection{Gradient updates}  
Several data structures are maintained during the SMO training~\cite{plattsmo}. An essential data
structure, $\gamma$, is described as follows:
\begin{equation}\label{eq:fcache}
\gamma_i = \mathsmaller\sum_{j}\alpha_j y_j \Phi(\bx_i) \cdot \Phi(\bx_j) - y_i
\end{equation}
The relationship between $\gamma$ and the gradient of~\eqref{eq:lagdual} is
shown in Table~\ref{table:model}. For the rest of the paper,
$\gamma$ and gradient are used interchangeably.
 In all the algorithms proposed in this work, the key
 component in the gradient of the dual objective
 function~\eqref{eq:lagdual}, $\gamma$, is maintained for all the samples in the
 training set/non-shrunk samples and
 not just the recently optimized samples at a given iteration for reasons
 explained in Section~\ref{sec:fcachesync}.

 The
 update equation is shown below:
     \begin{equation}\label{eq:fup}
     \begin{split}
     \gamma^{new}_{i} = \gamma^{old}_{i} + \\
     y_{up}*(\alpha^{new}_{up} - \alpha^{old}_{up})*(\Phi(x_{up}).\Phi(x_{i})) \\
     y_{low}*(\alpha^{new}_{low} - \alpha^{old}_{low})*(\Phi(x_{low}).\Phi(x_{i}))\\
     \end{split}
     \end{equation}
     where
     \begin{equation}\label{eq:setinfo}
     \begin{split}
     i\in I_0\cup I_1\cup I_2\cup I_3\cup I_4. \quad I_0=\{i: 0 < \alpha_i<C\},\\
     I_1=\{i:y_i=1, \alpha_i=0\}, I_2=\{i:y=-1, \alpha_i=C\}, \\
     I_3=\{i:y_i=1,\alpha_i=C\}, I_4=\{i:y_i=-1,\alpha_i=0\}
     \end{split}
     \end{equation}

\subsubsection{Working Set Selection}
\label{sec:wss}
The Working set selection describes the selection of samples to be
evaluated at each step of the algorithm. Since we work with the first derivative
of~\eqref{eq:lagdual} (refer to~\eqref{eq:bminmax}), this working set selection
        is addressed as first-order heuristics. Keerthi {\em et al.} have proposed multiple possibilities~\cite{keerthi}.
Algorithm one iterates over all examples in $I_0$ and the second
approach only evaluates
the \textit{worst KKT violators}, $\beta_{up}$ and $\beta_{low}$, where at each
step, they are calculated as shown in~\eqref{eq:bminmax}.
Of
these, we have adapted the second modification, and instead of having
two loops, we operate only in the innermost loop, avoiding the first
costly loop that examines all examples.  \par We do not compromise on
the accuracy of the solution $\bw$~\eqref{sol} because $1)$ we select
the pair of indices based on~\eqref{eq:bminmax} and not just using
$I_0\cup \{i_1, i_2\}$ as done in~\cite{keerthi} where $\{i_1,i_2\}$ is
a recently optimized pair and $2)$ of the nature of our $\gamma$ updates.
          \begin{equation}\label{eq:bminmax}
          \begin{split}
          \beta_{up}=min\{\gamma_i:i\in I_0\cup I_1\cup I_2\} \\
          \beta_{low}=max\{\gamma_i:i\in I_0\cup I_3\cup I_4\}
          \end{split}
          \end{equation}
These values are the two threshold parameters discussed in the optimized
version~\cite{keerthi}. The optimality condition for termination of the
algorithm (considering numerical issues) is 
\begin{equation}\label{eq:kktterm}
\beta_{up}+2*\epsilon \geq \beta_{low}
\end{equation}
where $\epsilon$ is a user-specified tolerance parameter. 

It can be seen from~\eqref{eq:bminmax},~\eqref{eq:fup} and~\eqref{eq:setinfo}, that the worst violators
are gathered by considering all the samples, not just the recently optimized
ones and the non-bound samples (i.e., $0<\alpha_k<C$) for the next iteration.

\subsubsection{Adaptive Elimination/Shrinking}
\label{sec:aes}
Shrinking is a mechanism to expedite the convergence of SVM training phase
by eliminating the samples, which would not contribute to the
hyperplane~\cite{joachims, libsvm}. 
With $I_1, I_2, I_3$ and $I_4$ defined as in~\eqref{eq:setinfo}, samples may be eliminated if they
    satisfy the following decision rule: 
    \begin{equation}\label{eq:shrink}
    \begin{split}
     i\in\{I_3\cup I_4\} \quad and\quad \gamma_i < \beta_{up} \\ or\\
     i\in\{I_1\cup I_2\}\quad and \quad \gamma_i > \beta_{low} 
    \end{split}
    \end{equation}
    This heuristic is explained in the Figure~\ref{fig:shrink_depict}. The eliminated
    samples belong to one of the two classes: $a)$ ones that have
    $\alpha=0$ and $b)$ those with
    $\alpha=C$. 
	

\subsection{Programming Models}
This paper uses two programming models - MPI~\cite{mpi1, mpi2} and
Global Arrays~\cite{nieplocha:psw94} for designing scalable SMO on
distributed systems. Due to space limitations, we provide a brief
background of Global Arrays and suggest other literature for
MPI~\cite{mpi1, mpi2}.
\subsubsection{Global Arrays}
The Global Arrays programming model provides abstractions for
distributed arrays, load/store semantics for local partition of the
distributed arrays, and one-sided communication to the remote
partitions.  Global Arrays leverages the communication primitives
provided by Communication Runtime for Exascale
(ComEx)~\cite{vishnu:cass13}.  Global Arrays programming model has been
used for designing many scalable applications in domains such as
chemistry~\cite{nwchem} and sub-surface modeling~\cite{STOMP:homepage}. 
The Global Arrays infrastructure is useful in storing the entire dataset
in a compressed row format. The easy access to local and remote portions
of distributed arrays facilitates a design of algorithms which would
need asynchronous read/write access to the arrays. Global Arrays uses
ComEx network communication layer for one-sided communication.

\section{Solution Space}
\label{sec:design}
\begin{table} 
		\centering 
		\caption{Representative notations used and their explanation} 
		\begin{tabular}{|c|c|} 
				\hline Name & Symbol\\ 
				\hline \# of Processors & $p$ \\  
				\hline \# of Training Points & $\mathcal{N}$ \\  
				\hline Class label & $y_{k}$ \\  
				\hline Lagrange multiplier & $\alpha_{k}$ \\  
				\hline Set of Support Vectors & $\zeta$ \\  
				\hline Working set& $\pi$ \\  
				\hline $\delta L_D/\delta\alpha_{k}$,
				$\gamma_{k}*y_{k}$~\eqref{eq:fcache} & $\nabla_{k}$ \\  
				\hline Hyperplane threshold&$\beta$\\
				\hline Sample in CSR form&$\breve{x}$\\
				\hline Indices set $I_{0-4}$
				in~\eqref{eq:setinfo}&$\varsigma$\\
				\hline User-Specified Tolerance & $\epsilon$\\
				\hline Avg $\langle,.~,\rangle$ time & $\lambda$\\
				\hline Row-Pointer Array & $\psi$\\
				\hline Average sample length
                                $\left|\bm{x_k}\right|$ & $ m$\\
				\hline Network Latency & $l$ \\
				\hline Network Bandwidth & $\frac{1}{G}$ \\
				\hline
		\end{tabular} 
		\label{table:model} 
\end{table}
This section begins with a presentation of various steps of the sequential
SVM training algorithm~\ref{algo:seqsmo}, which is followed by a discussion of the data
structures organization using the parallel programming models.
Section~\ref{sec:tct} introduces the parallel training algorithm 
of the Original algorithm, and presents its time-space complexity. 
This is a followed by a  discussion and analysis of multiple  parallel shrinking algorithms~\ref{sec:fcachesync}
\begin{algorithm}
\KwIn{$\mathcal{C}$, $\mathcal{\sigma}$, $\mathcal{X}\in\Re^{\mathcal{N}\times d}$, $y_i\in\{+1, -1\}, i=1, 2,\ldots \mathcal{N}$}
\KwData{$\bm{\alpha} \in \Re^{\mathcal{N}\times 1}$}
\KwResult{$\zeta$}
Initialize $\gamma_i=-y_i, \alpha_i=0, \forall i $\;
$i_{low}=\{j \mid y_j=1, j\in\{1,2,\ldots \mathcal{N}\}\}$\;
$i_{up}=\{k \mid y_k=-1, k\in \{1,2,\ldots \mathcal{N}\}\}$ \;
\Repeat{\eqref{eq:kktterm} succeeds}{
    Update
        $\alpha_{i_{low}}$ and $\alpha_{i_{up}}$ using~\eqref{eq:alphaup}\;
    Assign $i_{low}$ and $i_{up}$ to one of $\varsigma$ using~\eqref{eq:setinfo}\;
    $\forall i$, Update $\gamma_i$ using~\eqref{eq:fup}\;
    Calculate new $\beta_{low}$ and $\beta_{up}$ using~\eqref{eq:bminmax}\;
}
\caption{Improved SMO - Modification 2~\cite{keerthi}}
\label{algo:seqsmo}
\end{algorithm}

\begin{figure*}[htbp]
		\centering
		\begin{minipage}[t]{0.32\textwidth}
		\centering
				\includegraphics[width=\columnwidth]
                                {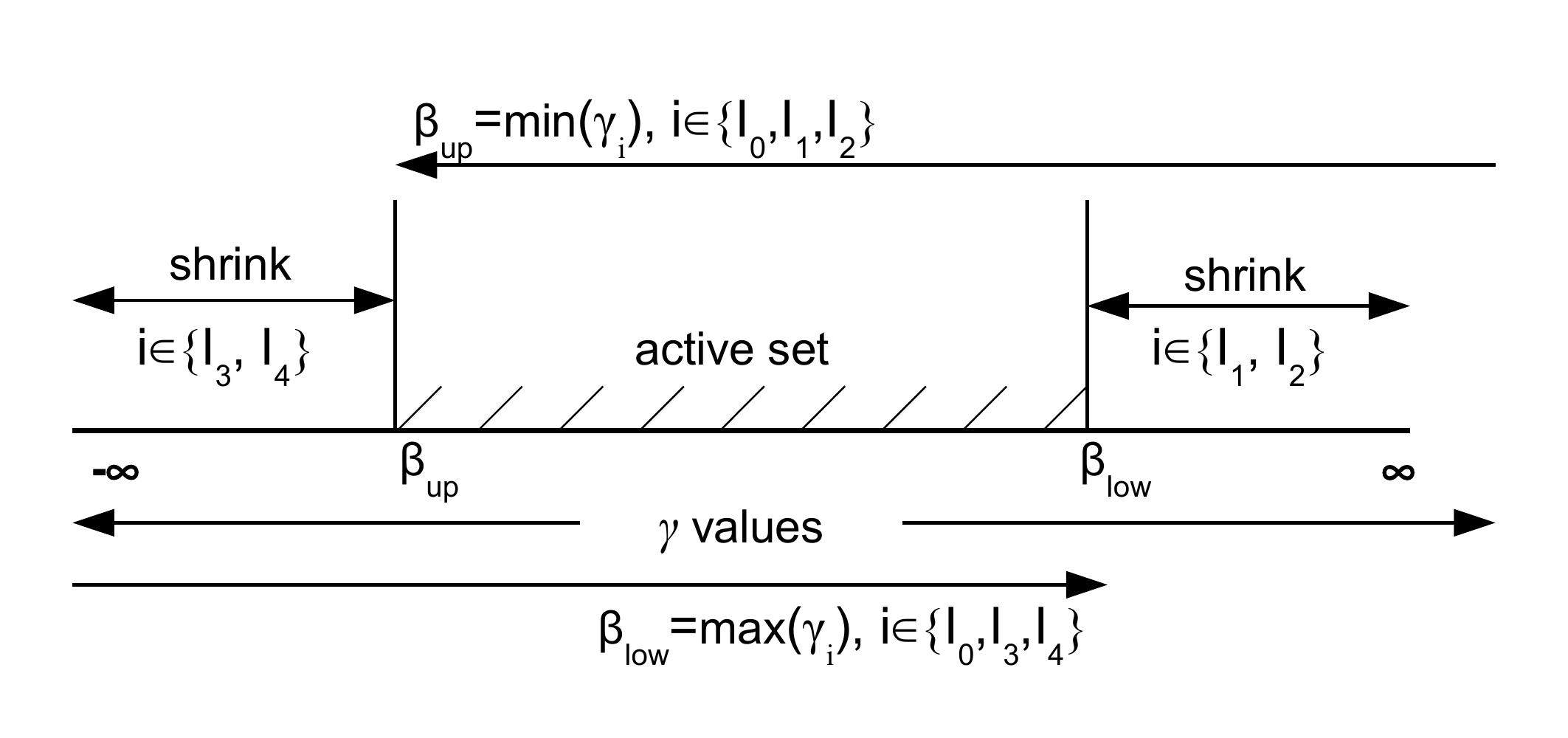}
				
				\subcaption{}
				\label{fig:shrink_depict}
		\end{minipage}
		\begin{minipage}[t]{0.28\textwidth}
				\centering
				\includegraphics[width=\columnwidth
                                ]{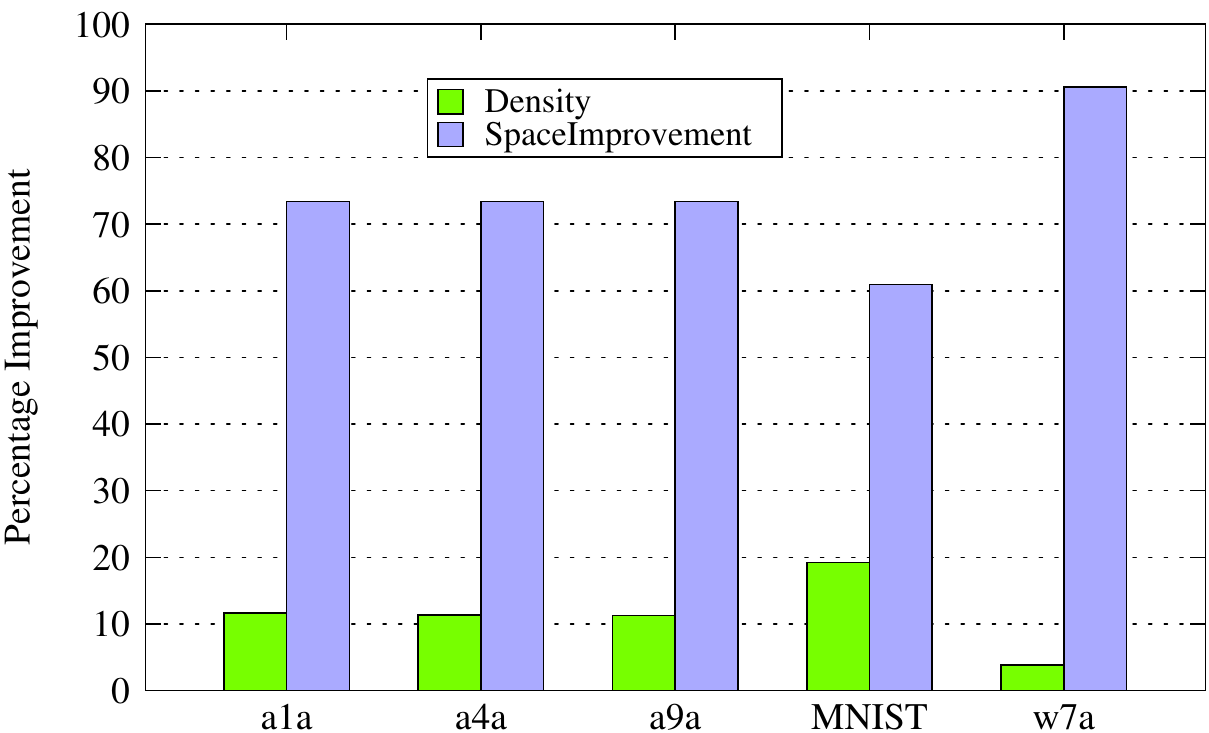}
                                \subcaption{}
				\label{fig:csr_adv}
		\end{minipage}
		\begin{minipage}[t]{0.39\textwidth}
				\centering
				\includegraphics[width=\columnwidth,
                                ]{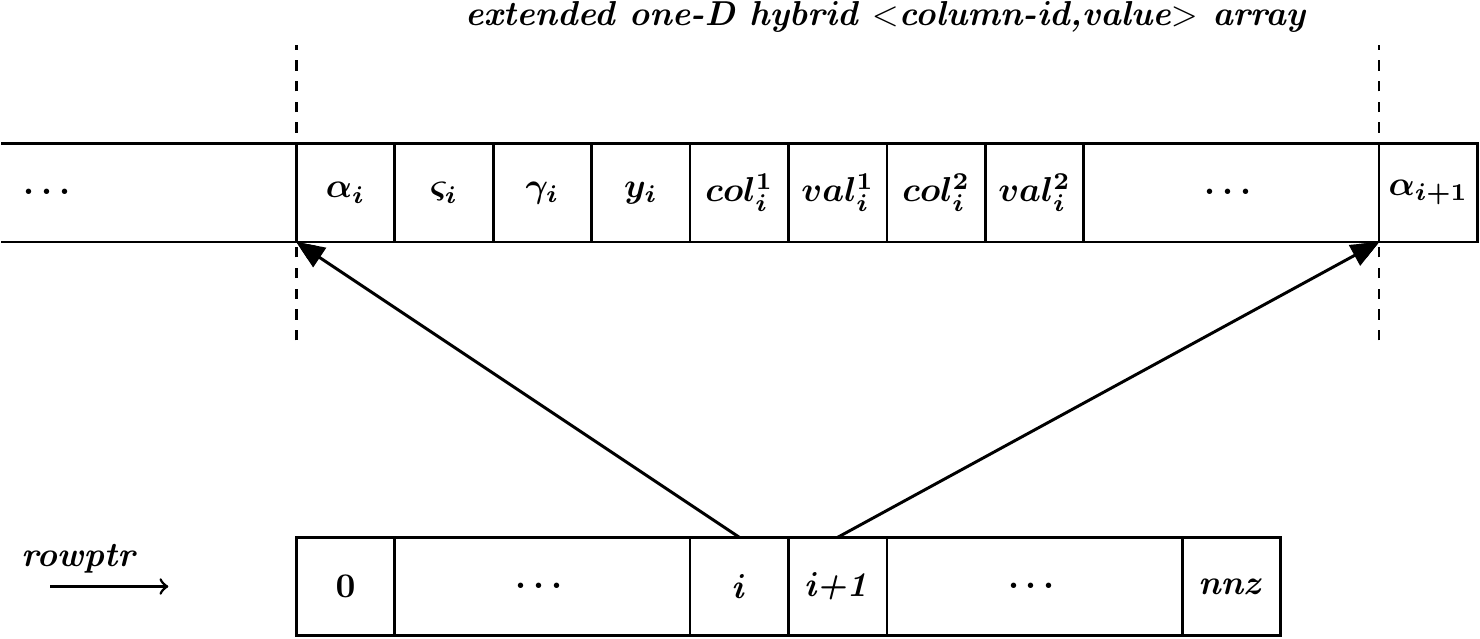}
                                \subcaption{}
                            \label{fig:exsamp} 
                        \end{minipage}
\caption{
    (\subref{fig:shrink_depict}): Shrunk ids when~\eqref{eq:kktterm} is not
                                    satisfied (non-optimality). 
        Refer to~\eqref{eq:shrink} for the shrinking criterion.
				(\subref{fig:csr_adv}): Memory Space Conservation with CSR
(\subref{fig:exsamp}): $\breve{x}$: An extended sample
                                    prototype in CSR representation. Refer to
                                        Table~\ref{table:model}
                                 for the meaning of the symbols.
      }
\end{figure*}

\subsection{Preliminaries}
Algorithm~\ref{algo:seqsmo} shows the key steps of our sequential SVM algorithm.
This is used as a basis for designing parallel SVM algorithms, with
(Algorithms~\ref{algo:parsmo_p2} and its variant) and
without (Algorithm~\ref{algo:parsmo_p1}) shrinking.
Using Table~\ref{table:model} as reference, at each iteration,  $\forall i$, $\alpha_i$ is
calculated based on~\eqref{eq:alphaup}. In most cases,  the objective function is positive
definite ($\rho < 0$)~\ref{eq:rho}, which is used as the basis for
update. An approach proposed by Platt {\em et al.}~\cite{plattsmo} can be used for the update equations, when 
$\rho>0$. 

\begin{equation}\label{eq:alphaup}
\begin{split}
        \alpha_{i_{low}}^{new}= \alpha_{i_{low}}-
        y_{i_{low}}*(\gamma_{i_{up}}-\gamma_{i_{low}})/\rho 
        \\
        \alpha_{i_{up}}^{new}= \alpha_{i_{up}}+
y_{i_{low}}*y_{i_{up}}*(\alpha_{i_{low}}-\alpha_{i_{low}}^{new}) 
    \end{split}
        \end{equation}
        where
        \begin{equation}\label{eq:rho}
        \begin{split}
    \rho=2*\Phi(\bx_{i_{low}})\cdot\Phi(\bx_{i_{up}})\\
        -\Phi(\bx_{i_{up}})\cdot\Phi(\bx_{i_{up}})
            -\Phi(\bx_{i_{low}})\cdot\Phi(\bx_{i_{low}})
    \end{split}
    \end{equation}
Once~\eqref{eq:kktterm} is satisfied, $\beta$ in~\eqref{eq:classi} is calculated
as:
\begin{displaymath}
\beta=\left\{ \begin{array}{ll}
\sum_{i\in I_0}\gamma_i/\left|I_0\right| & \textrm{if $\left|I_0\right|\neq 0$
}\\
(\beta_{low}+\beta_{up})/2 & \textrm{otherwise}
\end{array}
\right.
\end{displaymath}

\subsubsection{Distributed Data Structures}
There are several data structures required by algorithms~\ref{algo:parsmo_p1} and~\ref{algo:parsmo_p2} 
, which need to be distributed across different compute
nodes. These data structures include the $\mathcal{X}$ for the input
dataset, $y$ for the sample label, $\bm{\alpha}$ for the Lagrange
multipliers, $\gamma$ and $\varsigma$.  As presented earlier, the
computation can be re-formulated to using a series of kernel
calculations. The individual kernel calculations may be stored in a
kernel cache, which itself can be distributed among different
processes. 

However, there are several reasons for avoiding kernel cache for large
scale systems. The space complexity of complete kernel cache is
$\Theta(\mathcal{N}^2)$, which is prohibitive for large inputs - a
primary target of this paper. At the same time, the temporal/spatial
reuse of individual rows of kernel cache is low as the $i_{up}$ and
$i_{low}$ typically do not exhibit a temporal/spatial pattern.  As the
architectural trends exhibit, the available memory per computation unit
(such as a core in a multi-core unit) is decreasing rapidly, and it is
expected that simple compute units such as Intel Xeon Phi would be
commonplace~\cite{exascale:report08}, while Graphics Processors are
already ubiquitous.  At the same time, these compute units provide
hardware support for wide-vector instructions, such as fused
multiply-add. As a result, the cost of
recomputation is expected to be much lower than either caching the
complete kernel matrix or conducting off-chip/off-node data movement to
get the individual rows of kernel matrix distributed across multiple
nodes. Hence, the proposed approaches in this paper avoid the kernel
cache altogether.

\subsubsection{Data Structure Organization}
\label{sec:dso}
The organization of distributed data structures plays a critical role in
reducing time and space complexity of algorithm~\ref{algo:seqsmo}. 
Most datasets are sparse in nature, with several datasets having less than
20\% density.
Figure~\ref{fig:csr_adv} shows the percentage of memory space
conserved when using  a compressed sparse row (CSR)~\cite{dongarra} representation. As
shown in Figure~\ref{fig:exsamp}, co-locating the algorithmic related data
structures and making the
column indices as part of representation makes little change in the
density and the reduction in space complexity outweighs the additional
bookkeeping for the boundaries of various samples.

The core steps of computation requires several kernel calculations and
frequent access to the data structures such as the $y$, $\bm{\alpha}$,
and $\gamma$. Among these, $y$ is a read-only data structure, while other
data structures are read-write. The organization of these data
structures with $\mathcal{X}$ has a significant potential of improving
the cache-hit rate of the system, by leveraging the spatial locality.
Although a few of these data structures are read-write, the write-back
nature of the caches on modern systems make them a better design choice
in comparison to the individual data structures distributed across
multiple processes in the job. An additional advantage of co-location of
these data structures with $\mathcal{X}$ is that load balancing among
processes is feasible which requires contiguous data movement of
samples, instead of several individual data structures.

For the proposed approaches,  CSR is implemented using Global
Arrays~\cite{nieplocha:psw94} programming model.
Global Arrays provides semantics for collective creation of a compressed
row, facilitating productive use of PGAS models for
algorithms~\ref{algo:parsmo_p1} and \ref{algo:parsmo_p2}.
GA provides Remote Memory Access semantics for
traditional Ethernet based interconnects and Remote Direct Memory
Semantics (RDMA), making it effective for distributed systems such as
based on Cloud and tightly-connected supercomputers.

\begin{algorithm}
\KwIn{Samples $\breve{x}$ and $\breve{y}$, len($\breve{x}$), len($\breve{y}$), where len represents the number of cells in the representation}
\KwOut{$\langle\breve{x},\breve{y}\rangle$, the inner product of $\breve{x}$ and $\breve{y}$}
\tcc*[h]{shift past the padded data}\;$\grave{x}\gets \breve{x}+4$, $\grave{y}\gets \breve{y}+4$\;
$s1\gets 0$, $s2\gets 0$, $dp\gets 0$\;
\While {$s1 < len(\grave{x})-4~\&~s2<len(\grave{y})-4$}
{
    \uIf{$|\grave{x}[s1]-\grave{y}[s2]|<0 $ }
    {
        $dp\gets dp+\grave{x}[s1+1]*\grave{y}[s2+1]$\;
        $s1 \gets s1 + 2 $\;
        $s2 \gets s2 + 2 $\;
    }
    \uElseIf{$ \grave{x}[s1]<\grave{y}[s2]$}
    {
        $s1 \gets s1 + 2 $\;
    }
    \Else
    {
        $s2 \gets s2 + 2 $\;
    }
}
\Return{$\langle\breve{x},\breve{y}\rangle$}\;
\caption{Inner Product using CSR format}
\label{algo:dp}
\end{algorithm}

Algorithm~\ref{algo:dp}
shows the pseudocode for the inner product calculation - the most
frequently executed portion in our implementations. While the CSR representation is not
conducive for using hardware based vector instructions, we leave the use
of such vector units as a future work. The primary objective in this
paper is to minimize the space complexity by using CSR representation
not used by other papers such as~\cite{fsv}.
The inner product is used as a constituent value
in $\gamma$ (line~\ref{algo:fup} in algorithm~\ref{algo:parsmo_p1})
calculation by using a simple linear algebra trick.

\begin{algorithm}
\DontPrintSemicolon
\KwData{$p$: \# processors, $P_q$: $q$-th processor, $0\leq q<p$,\\ 
$i\in \left[\, q*|\mathcal{X}|/p,~(q+1)*|\mathcal{X}|/p \,\right)$
}
\KwIn{$\mathcal{C}$, $\mathcal{\sigma}$,
$\hat{\mathcal{X}}\in\Re^{\frac{\mathcal{N}}{p}\times d}$, $y_i\in\{+1, -1\}, \forall
i$,\\ $\bm{\alpha} \in \Re^{\frac{\mathcal{N}}{p}\times 1}$}
\SetKwFunction{GAget}{GAget}
\SetKwFunction{GAput}{GAput}
\KwResult{$\zeta$}
Initialize $\gamma_i=-y_i, \alpha_i=0, \forall i $, $i_{low}, i_{up}$\;
\Repeat{no \textit{KKT} violators}
{
            \tcp*[h]{MPI Broadcast Operation}\;
    Receive $\bm{x}_{i_{low}}$, $\bm{x}_{i_{up}}$ from proc$\#0$\;\label{algo:1}
    Update
        $\alpha_{i_{low}}$ and $\alpha_{i_{up}}$ using~\eqref{eq:alphaup}\;
    Constrain ${\alpha}s$ as per~\eqref{eqconst}\;\label{algo:2}
    \For{$\forall$i}
    {
		$li\leftarrow \psi_{i}$\;
		$hi\leftarrow \psi_{i+1}-1$\;
                $\breve{x}\leftarrow \GAget(li, hi)$\;
            \tcc*[h]{$\breve{x}[2] := Gradient$. Refer to Figure~\ref{fig:exsamp}}\;
            Update $\breve{x}[2]$ using~\eqref{eq:fup}\;\label{algo:fup}
        \If{$i==i_{up}~or~i==i_{low}$}
        {
            $\breve{x}[0]\leftarrow\alpha_{i_{low}}~or~\alpha_{i_{up}}$\;
            Update $\varsigma$ using~\eqref{eq:setinfo}\;
            \tcp*[h]{global copy update}\;
        }
            $\GAput(\breve{x}[0:2])$\tcc*[r]{first 3 cells} \label{algo:gaput}
    }
	$\beta_{up, local} \leftarrow min(\beta_{up_{i}}) 
	$\;
	$\beta_{low, local} \leftarrow max(\beta_{low_{i}}) 	$\;
	\tcp*[h] {Using MPI Allreduction}\;
  $\beta_{up}$, $\beta_{low}  
  \leftarrow~GlobalMinMax(\beta_{up, local}, \beta_{low, local}, p)$\;
}
\caption{Algorithm~\ref{algo:seqsmo} parallel version; $q$-th CPU perspective.}
\label{algo:parsmo_p1}
\end{algorithm}


\begin{algorithm}
    \SetKwFunction{GAget}{GAget}
    \SetKwFunction{GAput}{GAput}

\For{$\forall i \in \pi_q$}
{
    $li\leftarrow\psi[i]$,$hi\leftarrow\psi[i+1]-1$\;
    $\breve{x}\leftarrow\GAget(li,hi)$\; 
    \If{$i == i_{up}~or~i_{low}$}
    {
        update $\breve{x}[0],~\breve{x}[1]$\tcc*[r]{$\alpha, \varsigma$}
    }
    update $\breve{x}[2]$\tcc*[r]{$\gamma$}
    $\GAput(\breve{x}[0:2])$\;
    \If{$\neg shrinkitercounter$}
    {
       apply~\eqref{eq:shrink} to $\breve{x}$\;
       update $\pi_{q}$\; 
    }
}
\eIf{$\neg shrinkitercounter$}
{
    $shrinkitercounter\leftarrow MPI\_Allreduce(\left|\pi_{q}\right|)$\;
}
{
    $shrinkitercounter{-}{-}$\;
}
update global $\beta$ values\;

\caption{Parallel update of data structures following shrinking, $q-$th CPU
    perspective.}
\label{algo:dsup}
\end{algorithm}

\begin{algorithm}
\DontPrintSemicolon
\KwData{$p$: \# processors, $P_q$: $q$-th processor, $0\leq q<p$, 
 $\pi$}
\KwIn{$\mathcal{C}$, $\mathcal{\sigma}$,
$\hat{\mathcal{X}}\in\Re^{\frac{\mathcal{N}}{p}\times d}$, $y_i\in\{+1, -1\}, \forall
i$,\\ $\bm{\alpha} \in \Re^{\frac{\mathcal{N}}{p}\times 1}$}
\SetKwData{Shrink}{shrink}
\SetKwData{msc}{min\_shrink\_counter}
\SetKwFunction{GAget}{GAget}
\SetKwFunction{GAput}{GAput}
\SetKwFunction{Gradient}{gradientreconstruct}
\KwResult{$\zeta$}
$i\in \left[\, startindex,~endindex \,\right)$, where $startindex\gets q*|\mathcal{X}|/p$, 
$ endindex\gets (q+1)*|\mathcal{X}|/p$\;
Initialize $\gamma_i\gets -y_i, \alpha_i\gets 0, \forall i $, $i_{low}, i_{up}$\;
$\pi_{q}$[$0\ldots(endindex-startindex)]\gets startindex\text{--}endindex$\;
\While{$1$}
{
    $outloop\gets false$, $tolflag\gets false$\;
    \If{$shrink$}
    {
        \If{$\beta_{up} < \beta_{low}-20 \cdot \epsilon$} \protect\label{algo:p2:cond1}
        {
            \Repeat{failed}
            {
    Receive $\bm{x}_{i_{low}}$, $\bm{x}_{i_{up}}$ from proc$\#0$\;
    Update
        $\alpha_{i_{low}}$ and $\alpha_{i_{up}}$ using~\eqref{eq:alphaup}\;
    Constrain ${\alpha}s$ to satisfy~\eqref{eqconst}\;
                Perform Algorithm~\ref{algo:dsup}\;

            }
        }
        \Else
        {
            $tolflag\gets true$\;
        }
    }
    \Else(\tcc*[f]{shrinking done once})
    {
        \If{$\beta_{up} < \beta_{low}-2 \cdot \epsilon$} \protect\label{algo:p2:cond2}
        {
            \Repeat{failed}
            {
    Receive $\bm{x}_{i_{low}}$, $\bm{x}_{i_{up}}$ from proc$\#0$\;
    Update
        $\alpha_{i_{low}}$ and $\alpha_{i_{up}}$ using~\eqref{eq:alphaup}\;
    Clip ${\alpha}s$ to the box constraint~\eqref{eqconst}\;
                Call Algorithm~\ref{algo:dsup}\;
            }

        }
        \Else
        {
            $tolflag\gets true$\;
        }
    }
    \If{$\Shrink~\& \left(\,outloop \| tolflag\,\right)$}
    {
        \Gradient()\tcc*[r]{Algorithm \ref{algo:fcachereconst}}\
        \If {$\beta_{up} < \beta_{low}-2 \cdot \epsilon $}
        {
            $shrink\gets 0$\;
            $shrinkitercounter\gets \msc$\;
            Reset $\pi_{q}$\;

        }
        \Else
        {
            break\tcc*[r]{optimality reached}
        }
    }
    \Else
    {
        \If{$outloop\|tolflag$}
        {
        break\;
        }
    }
}
\caption{Parallel Shrinking with single call to
    Algorithm~\ref{algo:fcachereconst}. $q$-th CPU perspective.}
\label{algo:parsmo_p2}
\end{algorithm}

\subsection{Parallel SVM training Algorithms}
\label{sec:tct}
This section lays out the parallel algorithms -
with shrinking~\ref{algo:parsmo_p2} and without
shrinking~\ref{algo:parsmo_p1}. It also presents the reasoning behind
shrinking and the conditions which must be true at the point of
shrinking.

%
\subsubsection{Parallel algorithm}
Algorithm~\ref{algo:parsmo_p1} is a parallel 
\textit{\emph{no-shrinking}} variant of the 
sequential algorithm~\ref{algo:seqsmo}. This algorithm is also referred as {\em
    Original} in various sections of this paper. There are
    several steps in
this algorithm. Each process receives $\bm{x}_{low},\bm{x}_{up}$ from a default
process using the MPI broadcast primitive, which is a scalable
logarithmic operation in the number of processes.
 Each process independently calculates the new $\bm{\alpha}$
corresponding to $i_{up}$ and $i_{low}$. This results in a time
complexity of $O(l + m \cdot G) \cdot log(p)$ for network communication
and  three kernel calculations $3 \cdot \lambda$ (ignoring other integer
based calculation).

The {\em for-loop} over all samples for a process is the predominantly
expensive part of the calculation. Each iteration requires the
calculation of the $gradient$, which involves several kernel calculations,
the $\varsigma$ calculation and update of the global values of $\alpha_{up}$
or $\alpha_{low}$, if they are locally owned by the process.  The
\texttt{GAput} operation (line ~\ref{algo:gaput}) updates only the indices,
which were updated during the calculation, reducing the overall
communication cost. The computation cost of this step is $\Theta(\lambda
\cdot \frac{\left|\mathcal{X}\right|}{p})$. For a sufficiently large
$\frac{\left|\mathcal{X}\right|}{p}$, the $\varsigma$ calculation and the communication
cost to update the global copy of $\alpha$ can be ignored. The last step of
the algorithm is to obtain the globally maximum and minimum of
$\beta_{low}$, and $\beta_{up}$, respectively. This is designed using
MPI Allreduction operation which has a time-complexity of $\Theta(l
\cdot log(p))$ (The bandwidth term can be ignored, since this step
involves a communication of only two scalars).


 \subsection{Shrinking Algorithms}
            Joachims {\em et al.}~\cite{joachims} and Lin {\em et al.}~\cite{libsvm} have previously demonstrated
            the impact of adaptive elimination of samples - shrinking. This
            technique is a heuristic, since the sufficient conditions
            to identify the samples to be eliminated are unknown~\cite{joachims}. For the eliminated samples, the
            Lagrange multipliers are kept fixed and they are not considered
            during the working set selection and the check for optimality.
            This results in time-complexity reduction, since the gradient for eliminated samples is not computed. The primary
            intuition behind shrinking is that only a small subset of samples
            contributes towards hyperplane definition:
\begin{equation}\label{eq:shrinkcond1}
\begin{split}
                \mathcal{A}=\{\,k \mid
                \gamma_k<\beta_{low}~or~\gamma_k>\beta_{up}, 0<\alpha_k\,\} 
                \\
                    \textrm{and}\\
               \left|\mathcal{A}\right|\ll\left|\mathcal{X}\right|
                \end{split}
                \end{equation}
It is expected that when the optimization is at the early stage, some of the
bound samples $(\alpha_k =0,~\alpha_k=C)$ stabilize~\cite{joachims}.
                \begin{equation}\label{eq:shrinkcond2}
                \begin{split}
\textrm{At non-optimality after sufficient iterations:} \\
                \mathcal{\grave{A}}=\{\,k \mid
                \beta_{low}\geq\gamma_k\geq\beta_{up}\,\}
                \end{split}
                \end{equation}
                where $\mathcal{\grave{A}}$ is the set of violators from where working
                set variables are chosen and one or more samples from the set
                $\mathcal{X}-\mathcal{\grave{A}}$ can be eliminated without changing the
                current solution.
 Specifically,~\eqref{eq:shrink}
presents a variant of the condition proposed previously by Lin
{\em et al.} for shrinking. The overhead of calculating which samples to shrink
is expected to be $\Theta(1)$, since the computation only involves a few
conditions.

 However, there are several problems with this assumption. It is possible
 that samples with $\alpha\in\{0,C\}$ -  which were previously
 eliminated - eventually stabilize to a value between $0$ and $C$. A
 premature elimination of these samples may result in the incorrect
 definition of hyperplane. A {\em conservative} approach to decide on the
 execution of this condition may not be beneficial, since much of the
 calculation would likely have completed. In essence, it is very
 difficult to predict the point at which to execute this condition. Lin
 {\em et al.} have proposed to use $min(|\grave{A}|,1000)$ iterations as
 the point to perform shrinking. However, there is no intuitive
 reasoning behind selecting a value to begin or executed shrinking. 
 A discussion on spectrum of heuristics for shrinking is presented in
 the next section.
 \subsubsection{Shrinking Heuristics} 
\label{sec:sh}
The heuristics for shrinking considered in
this paper are to address the concerns of early elimination of the samples,
while still reducing the overall time for SVM convergence. In general, $\left|\zeta\right|$ $\ll$ $\left|\mathcal{X}\right|$. Using this {\em intuition}, we
propose several heuristics for shrinking, which are based on the
$\left|\mathcal{X}\right|$. An {\em aggressive} shrinking heuristic would use a
$n\cdot \left|\mathcal{X}\right|$ as the iteration count for initial shrinking,
where $n \ll 1$. A {\em conservative} shrinking heuristic could use a
larger
value of $n$.
This method is referred to as {\em numsamples} based approach
in Table~\ref{table:heuristics}.
An alternative technique is to use initial shrinking iteration counter to
be a {\em random} value, similar to the approach proposed by Lin {\em et al}.
For each of these heuristics, the subsequent calculation of shrinking
iteration is a minimum of $\left|\mathcal{X}\right|$ or
$|\grave{A}|$ (depending on the algorithms) and the value of shrinking
iteration calculated using the proposed heuristics. See
section~\ref{sec:heuristicsoverview} for further discussion on this topic.

\subsection{Gradient Reconstruction}
\label{sec:fcachesync}
Gradient reconstruction is an important step in ensuring that the
previously eliminated samples are not {\em false positives} and that they are on
the correct side of the hyperplane in the final solution.
Algorithm~\ref{algo:fcachereconst} shows the key steps involved in
updating $\gamma$ values during the gradient reconstruction step of the
algorithm~\ref{algo:parsmo_p2}. The algorithm~\ref{algo:parsmo_p2}
corresponds to shrinking with single gradient-reconstruction. An
algorithm, which corresponds to multiple gradient
reconstruction (Refer to Table~\ref{table:heuristics}) can be
derived from this. However, due to lack of space, it is not presented
explicitly.



Algorithm~\ref{algo:fcachereconst} finds the $\gamma$ values of all the eliminated 
samples from the previous gradient reconstruction. To achieve this, it
needs $\mathcal{X}-\grave{A}$, which results in the communication of samples
owned by each process. The time complexity of this step is $l +
|\mathcal{X}-\grave{A}| \cdot G$ $\approx$ $|\mathcal{X}-\grave{A}| \cdot G$. The
communication cost may be non-negligible for distributed systems, hence
it is necessary to consider heuristics which limit the execution of
gradient synchronization step. 
Also evident from the loop structure is the fact that the outer loop considers all eliminated
samples of the $q$-th CPU and updates their gradient values. This is a
computationally expensive operation since line~\ref{algo:freconst} involves kernel
calculations (\eqref{eq:fcache} from section~\ref{sec:smo}), so this
algorithm is called only when global violators are within a specific
threshold (e.g., lines~\ref{algo:p2:cond1} and~\ref{algo:p2:cond2} in
Algorithm~\ref{algo:parsmo_p2}). Since $\gamma$ plays an important role in both
the $\alpha$ updates~\eqref{eq:alphaup} and working set selection
(Section~\ref{sec:wss}), we maintain
it for all the active samples throughout the program execution. 

Considering a less-noisy dataset, $\left|\zeta\right|\ll\mathcal{N}$ and on
an average, $\pi_q = \frac{\zeta}{p}$. Then,
$\left|\omega\cap\zeta\right|$ is small if not $0$ and the computational
time complexity expected for $q$-th CPU for
Algorithm~\ref{algo:fcachereconst} is $
\left|\omega_q\right|\cdot\left|\zeta\right|\cdot\lambda =
\left|\frac{\mathcal{X}-\zeta}{p}\right|\cdot\left|\zeta\right|\cdot\lambda
$. The tradeoff between $\left|\omega\right|$ and  $\left|\zeta\right|$
is clear making this essential algorithm a bottleneck in achieving the
overall speedup in convergence. As a result, we have considered single
and multi heuristics for $\gamma$-reconstruction as shown in
Figure~\ref{table:heuristics}.

\begin{algorithm}
\KwData{$p$: \# processors, $P_q$: $q$-th processor, $0\leq q<p$, $\pi$}
\KwIn{$\mathcal{\sigma}$,
$\hat{\mathcal{X}}\in\Re^{\frac{\mathcal{N}}{p}\times d}$, $y_i\in\{+1, -1\}, \forall
i$}
\SetKwFunction{GAget}{GAget}
\SetKwFunction{GAput}{GAput}
\tcp*[h]{Gather eliminated samples of this process}\;
$li \leftarrow \psi_{0}$\;
$hi \leftarrow \psi_{|\mathcal{X}|}$\;
	\texttt{GAget}($li, hi$)\;
$\omega_q=\hat{\mathcal{X}}-\pi_q$\;
\For{$\forall \breve{a}\in \omega_q$}
{
    $my\gamma\leftarrow0$\;
    \For{$\{ \,\forall \breve{b}\in \mathcal{X} \mid \breve{b}[0] > 0 \,\}$}
    {
        $my\gamma~+=\breve{b}[0]*\breve{b}[3]*(\Phi(\ba)\cdot\Phi(\bb))$\;\label{algo:freconst}
    }
    $\breve{a}[2]=my\gamma - \breve{a}[3]$\;
	\texttt{GAPut}($\breve{a}[2]$)\;
}
\tcp*[h]{MPI All reduction}\;
Update global $\beta_{low}$ and $\beta_{up}$\;
\caption{Gradient Reconstruction; $q$-th CPU perspective.}
\label{algo:fcachereconst}
\end{algorithm}

\section{Empirical Evaluation}
\label{sec:perf}
This section provides an empirical evaluation of the proposed approaches
in the previous section.
The empirical evaluation is conducted across multiple dimensions:
datasets, number of processes, shrinking/no-shrinking, heuristics for
selection of shrinking steps. The performance evaluation uses up to 512
processes (32 compute nodes), and several datasets use between 1 and 32
compute nodes. As a result, the proposed approaches can be used on
multi-core machines such as a desktop, supercomputers or
cloud computing systems. For each dataset, we compare
our results with \texttt{LIBSVM}~\cite{libsvm}, version
3.17, with {\em shrinking enabled}.

The upcoming sections provide a brief description of the datasets,
experimental testbed and followed by empirical results. Due to
accessibility limitations, the performance evaluation is conducted on a
tightly connected supercomputer, although the generality of our proposed
solution makes it effective for cloud computing systems as well.

\begin{table}
		\centering 
		\caption{Dataset Characteristics and hyperparameter settings} 
		\begin{tabular}{|c|c|c|c|c|c|c|} 
				\hline
				Name & Training Set Size& Testing Set
                                Size&C&$\sigma^2$\\ 
				\hline
				MNIST&60000&10000& 10  & 25 \\ 
				Adult-7 (a7a) &16100&16461&32 & 64 \\ 
				Adult-9 (a9a) &32561&16281 &32 & 64 \\ 
				USPS &7291&2007 & 8 & 16 \\
				Mushrooms &8124&N/A & 8 & 64 \\ 
				Web (w7a) &24692&25057 & 32 & 64 \\ 
				IJCNN &49990&91701 & 0.5 & 1 \\ 
				\hline
		\end{tabular}
		\label{tab:datasets}
\end{table}

\subsection{Datasets}
Table~\ref{tab:datasets} provides a description of the datasets used for
performance evaluation in this paper.
The MNIST\footnote{\url{deeplearning.net/data}} dataset represents
images of handwritten digits. The dimensions are formed by flattening
the 28x28 pixel box into one-dimensional array of floating point values
between 0 and 1, with 0 representing black and 1 white. The 
10-class dataset is converted into a two-class one by representing even digits
as class -1 and odd digits as +1.  
The sparse binary Adult dataset represents the collected census data for
income prediction. Web dataset is used to categorize web pages based on their
text~\cite{plattsmo}. USPS represents a collection of handwritten text
recognition, collected by United States Postal Service.  The mushrooms
data set includes descriptions of hypothetical samples corresponding to
23 species of gilled mushrooms in the Agaricus and Lepiota Family. 
The IJCNN dataset represents
the first problem of International Joint Conference on Neural Nets
challenge 2001.
Hyperparameter settings for the datasets have been selected after doing
multi-fold cross-validation~\cite{libsvm}. These are shown in
Table~\ref{tab:datasets}. The hyperparameter $C$ is described in
section~\ref{sec:train} and $\sigma^2$ is the kernel width in the
Gaussian kernel: $K(\bx,\by)=\exp(-\|\bx-\by\|^2/2\sigma^2 )$. It is
straightforward to use other kernels in this work.

\subsection{Experimental Testbed}
All our experiments were run on PNNL Institutional Computing (PIC)
cluster~\footnote{\url{pic.pnnl.gov/resources.stm}}.  PIC Cluster
consists of 692 dual-socket nodes with 16 cores per socket AMD
Interlagos processors, running at 2.1 GHz with 64 GB of 1600 MHz memory
per node (2 GB/socket). The nodes are connected using InfiniBand QDR
interconnection network. The empirical evaluation consists of a mix of
results on single node (multi-core) and multiple nodes (distributed system). While the performance
evaluation is on a tightly-connected system, most modern Cloud providers
provide programming models such as MPI and Global Arrays, hence this
solution is deployable on them as well.

\subsection{Heuristics: An overview}
\label{sec:heuristicsoverview}
Table~\ref{table:heuristics} provides a list of heuristics, which are
used for evaluation on datasets presented in the previous section.
Specific values to {\em aggressive}, {\em conservative} and {\em average}
methods for shrinking are provided. To emulate the heuristics evaluated
by Lin {\em et al.}, we also compare several values of {\em random}
sample elimination as suggested in the Table.
Line $2$ in the table is to be interpreted as shrinking every 2
iterations({\em aggressive}), with
a single call to gradient reconstruction. Optimization proceeds without
shrinking after this call. Similarly, line $13$ can be read as shrinking
whenever the number of iterations reach half the number of samples
({\em conservative}) with
multiple calls to $\gamma$ reconstruction as deemed fit and optimization
proceeds with shrinking throughout until convergence.
\begin{table} 
		\centering 
		\caption{Heuristics. Description and classification. $\star$: Aggressive shrinking class,
                    $\bullet$: Conservative, $\diamond$: Average} 
		\begin{tabular}{|c|c|c|c|c|} 
				\hline\#& Shrinking Type & $\gamma$-Recon. &
                                Name&Class\\ 
				\hline 
			    $1)$ &None & N/A & Original &N/A\\
			    $2)$&random: 2 & Single & Single2&$\star$ \\
			    $3)$&random: 500 & Single & Single500&$\star$ \\
			    $4)$&random: 1000 & Single & Single1000&$\diamond$ \\
			    $5)$&numsamples: 5\% & Single & Single5pc&$\star$ \\
			    $6)$&numsamples: 10\% & Single &
                            Single10pc&$\diamond$ \\
			    $7)$&numsamples: 50\% & Single & Single50pc&$\bullet$ \\
			    $8)$&random: 2 & Multi & Multi2 & $\star$\\
			    $9)$&random: 500 & Multi & Multi500&$\star $ \\
			    $10)$&random: 1000 & Multi & Multi1000&$\diamond$ \\
			    $11)$&numsamples: 5\% & Multi & Multi5pc&$\star $ \\
			    $12)$&numsamples: 10\% & Multi & Multi10pc&$\diamond $ \\
			    $13)$&numsamples: 50\% & Multi & Multi50pc&$\bullet $ \\
				$14)$&Default  & Default & \texttt{LIBSVM}& N/A  \\
				\hline
		\end{tabular} 
		\label{table:heuristics} 
\end{table}

\begin{figure*}[htbp]
\begin{minipage}[t]{\columnwidth}
\centering
\includegraphics[width=0.8\columnwidth]{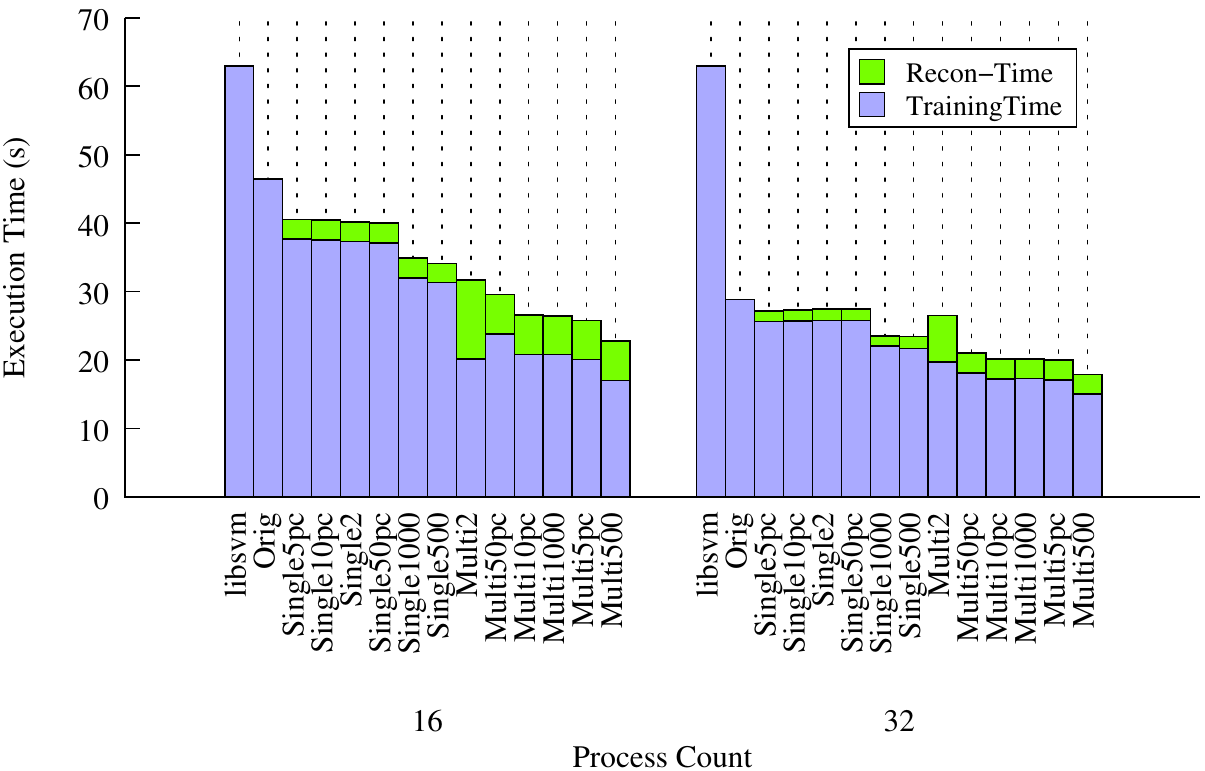}
\caption{Adult (a7a) Dataset Performance}
\label{fig:a7a}
\end{minipage}
\centering
\begin{minipage}[t]{\columnwidth}
\centering
\includegraphics[width=0.8\columnwidth]{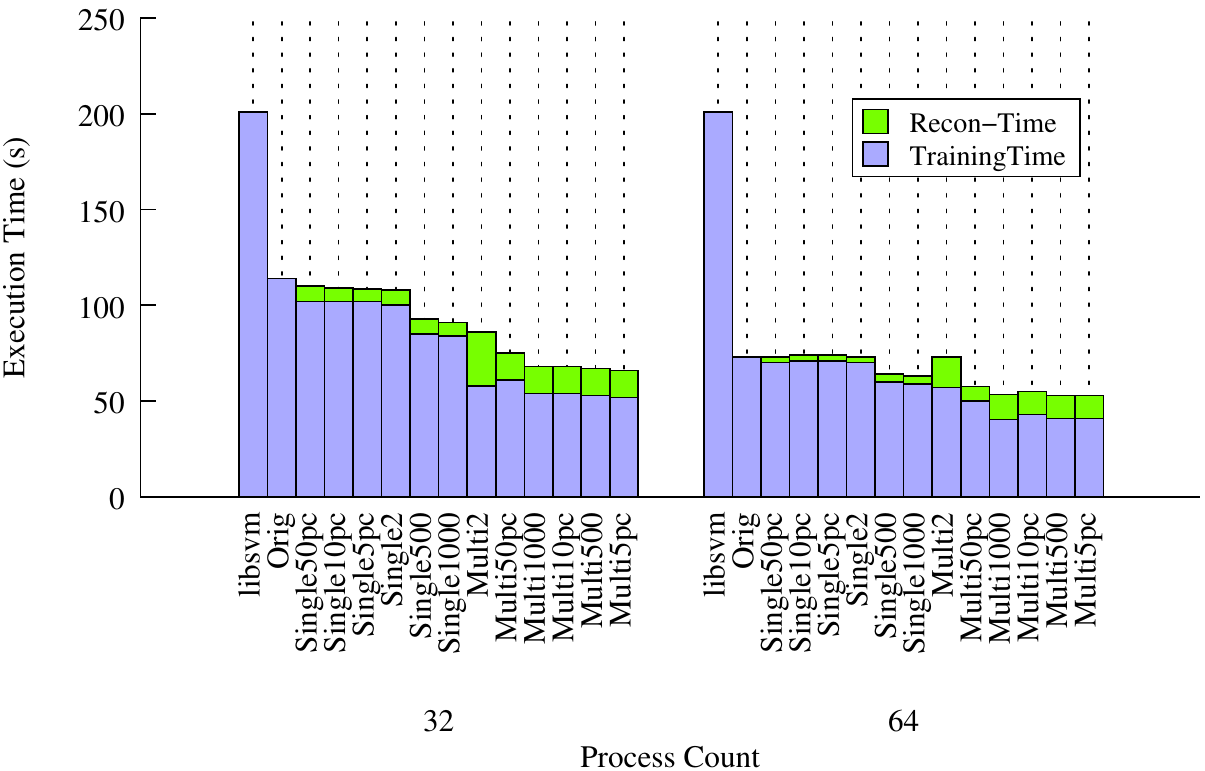}
\caption{Adult (a9a) Dataset Performance}
\label{fig:a9a}
\end{minipage}
\end{figure*}

\begin{figure*}[htbp]
		\centering
		\begin{minipage}[t]{\columnwidth}
				\centering
				\includegraphics[width=0.8\columnwidth]{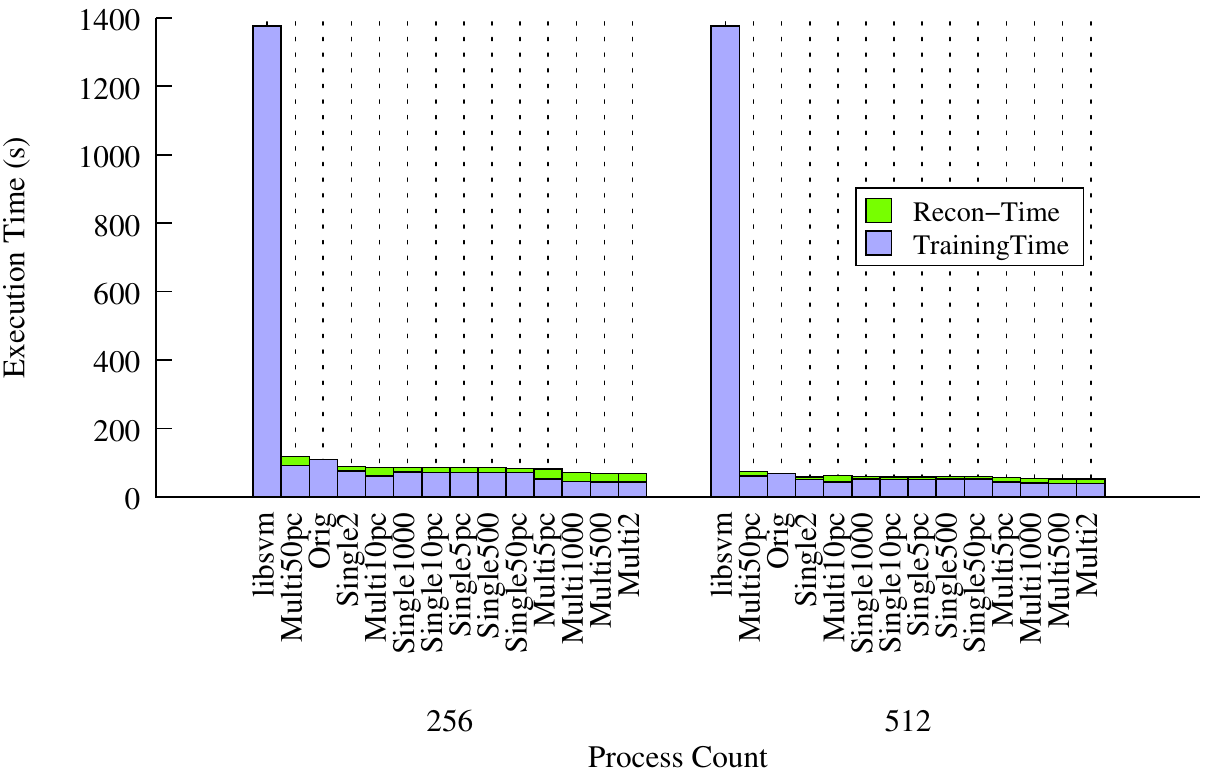}
				\caption{MNIST Dataset Performance }
				\label{fig:mnist}
		\end{minipage}
		\begin{minipage}[t]{\columnwidth}
				\centering
				\includegraphics[width=0.8\columnwidth]{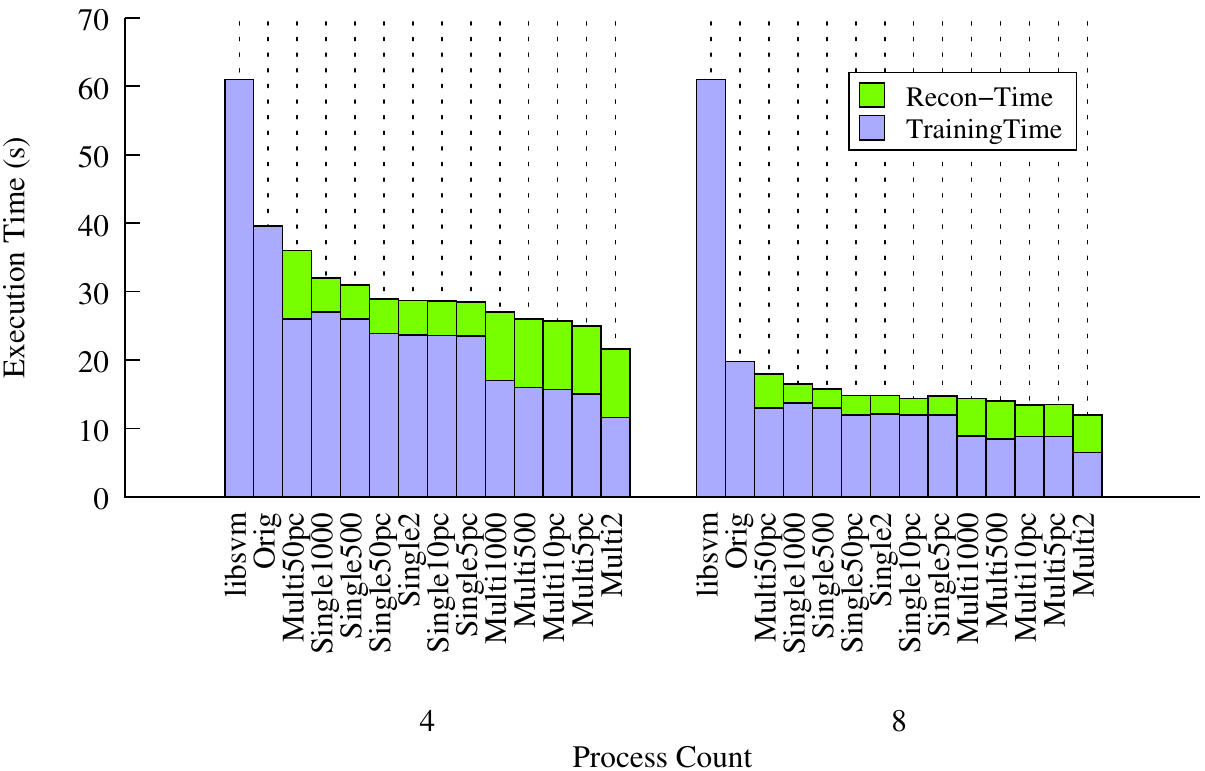}
				\caption{USPS Dataset Performance}
				\label{fig:usps}
		\end{minipage}
\end{figure*}

\begin{figure*}[htbp]
		\centering
		\begin{minipage}[t]{\columnwidth}
				\centering
				\includegraphics[width=0.8\columnwidth]{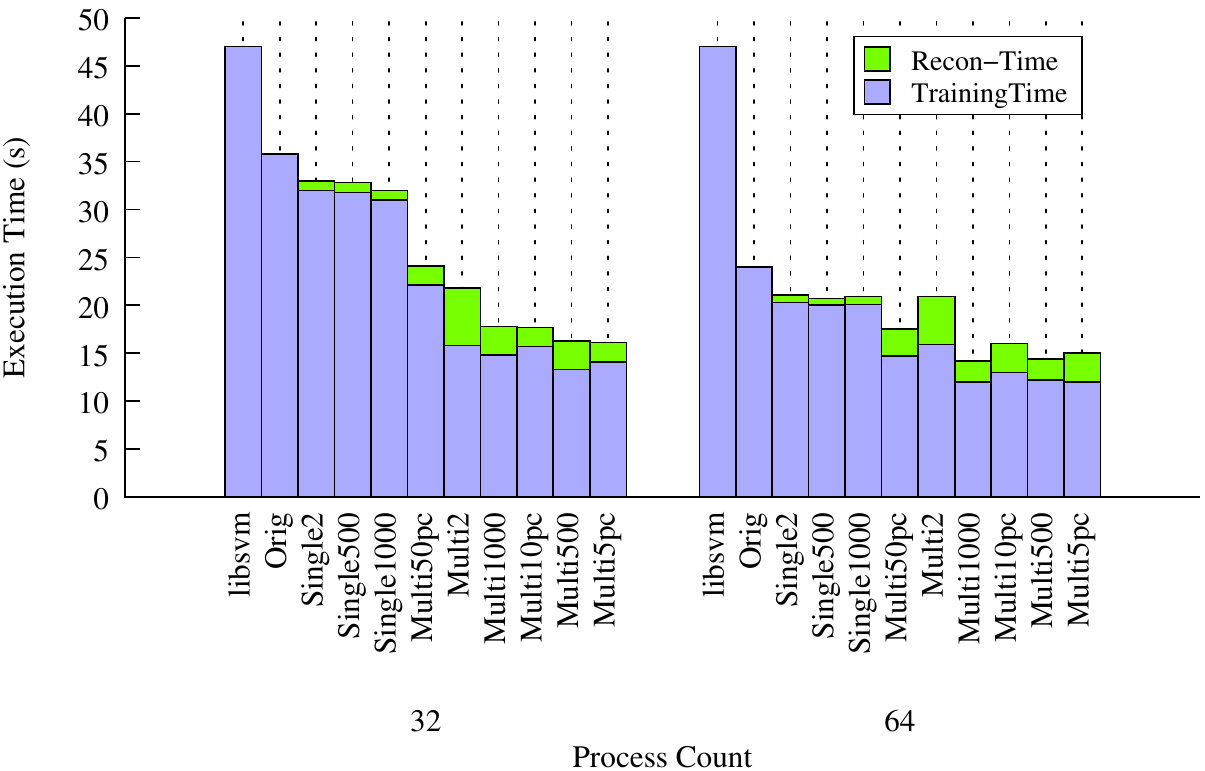}
				\caption{w7a Dataset Performance}
				\label{fig:w7a}
		\end{minipage}
		\begin{minipage}[t]{\columnwidth}
				\centering
				\includegraphics[width=0.8\columnwidth]{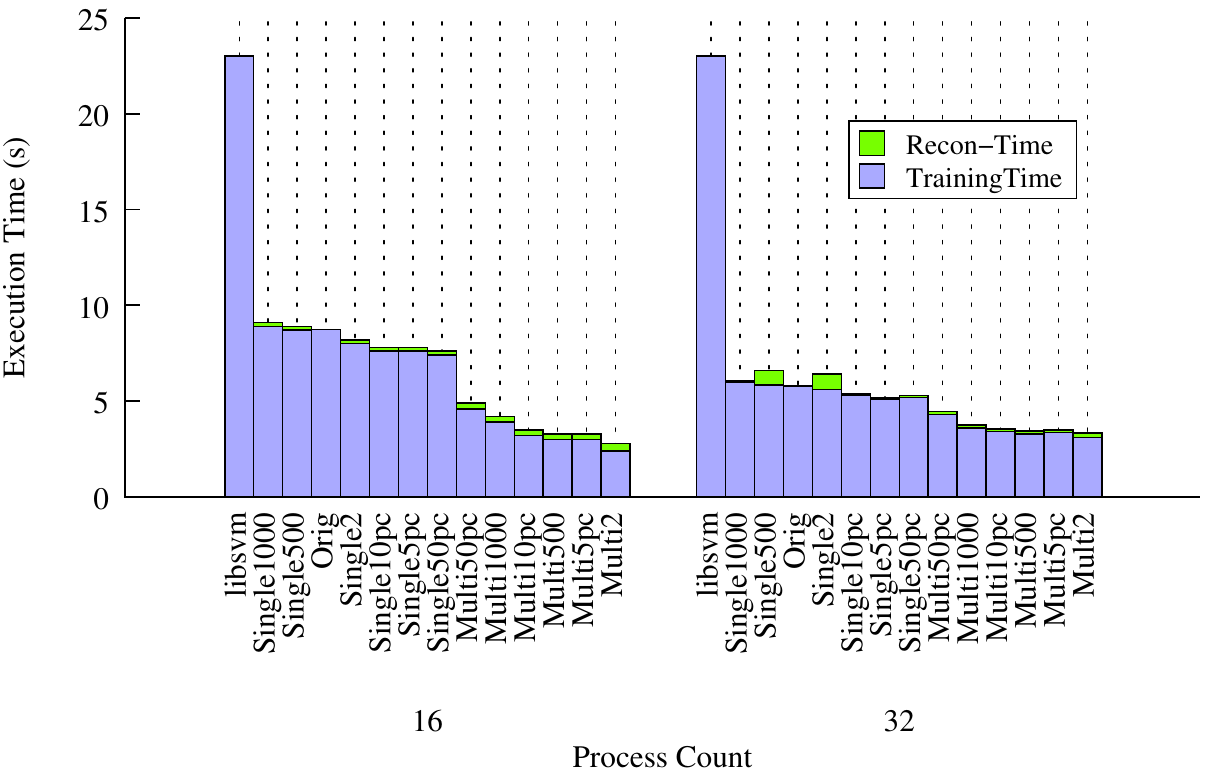}
				\caption{Mushroom Dataset Performance}
				\label{fig:mushrooms}
		\end{minipage}
\end{figure*}
\begin{figure*}[htbp]
\centering
		\begin{minipage}[t]{\columnwidth} 
                                \vspace{0pt}
				\centering
				\includegraphics[width=0.7\columnwidth]{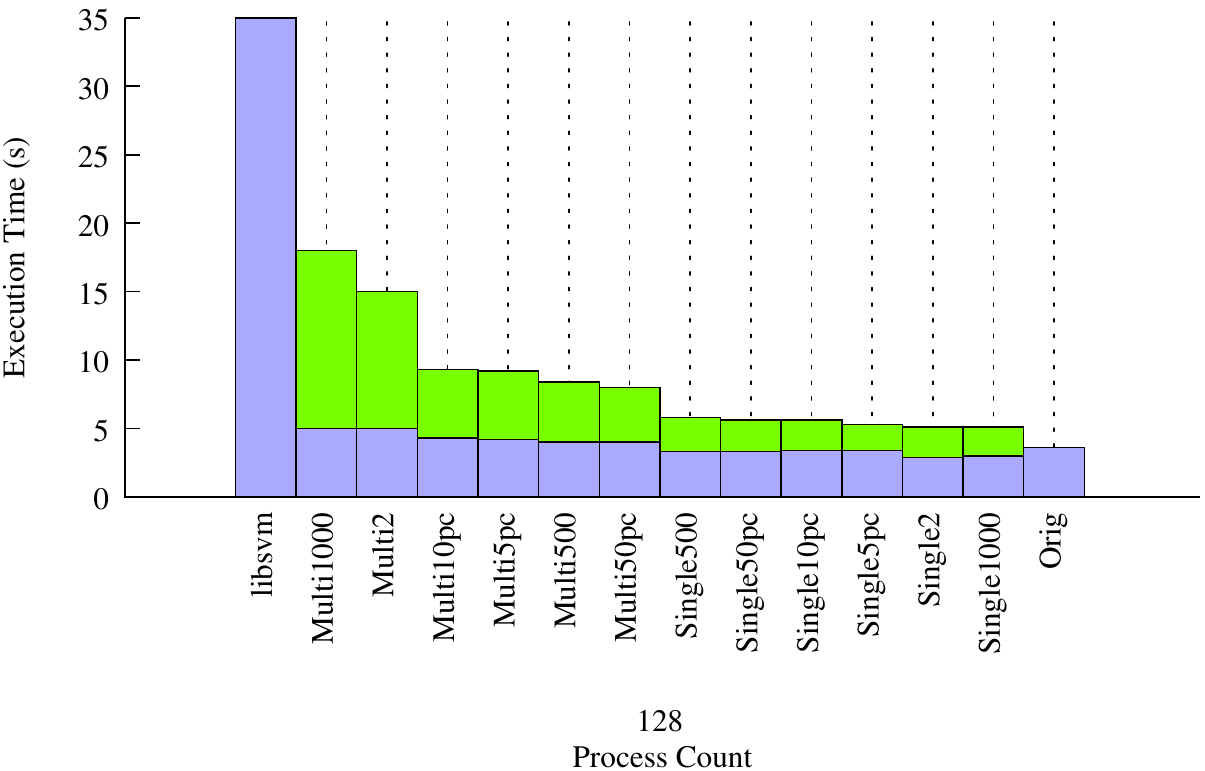}
				\caption{IJCNN Dataset
    Performance.}
				\label{fig:ijcnn1}
		\end{minipage}
               \hfill 
			   \centering
		\begin{minipage}[t]{\columnwidth}
                \vspace{0pt}
		\centering
                \resizebox{\columnwidth}{!}{
		\begin{tabular}{|c|c|c|c|c|c||} 
				\hline
                                    \hline
									Name & Test Acc. - Ours(\%) & Test Acc.-\texttt{LIBSVM}(\%) 
                                 & Speedup(Original) &
								 Speedup (\texttt{LIBSVM})\\ 
				\hline
				Adult-7 &83.75& 84.81 & 1.6x & 3.5x \\ 
				Adult-9 &84.68& 83.12 & 1.4x & 3.7x\\ 
				USPS &97.6&97.75 &1.7x &  5x \\ 
				Web &98.86&98.8 &1.6x & 3.3x \\ 
				MNIST&96.63 &  98.62 & 1.2x& 26x \\
                                Mushrooms& N/A&N/A&3x&8.2x\\
				IJCNN &90.3& 96.58 & N/A & 9x (Orig)\\ 
				\hline
		\end{tabular}
                }
		\caption{Summary of Results (Testing Accuracy and Relative
                Speedups between our best performing heuristic with the Original
                Implementation~\ref{algo:parsmo_p1} and \texttt{LIBSVM})} 
		\label{tab:testacc}
		\end{minipage}
\end{figure*}
\subsection{Results and Analysis}
Figures~\ref{fig:a7a} and~\ref{fig:a9a} show the results for Adult-7 and
Adult-9 datasets, respectively. A speedup of $\approx$ 2x is observed on
Adult-7 dataset using the Multi500 and Multi5pc heuristics in comparison
to Original algorithm, and 3-3.5x in comparison to \texttt{LIBSVM}. Among all
the approaches, Multi2 has the highest time in $\gamma$ Reconstruction
(referred as Recon-Time in the figures),
largely because it eliminates samples prematurely, while other
heuristics allow the $\alpha$ values to stabilize before elimination. It
is worthwhile noting that each of the Multi* heuristics are
better than Single shrinking for these datasets. For each of the adult
datasets, $\left|\zeta\right|$ $\ll$ $X$, which is a suitable condition for shrinking.
Since the proposed heuristics are precise, the accuracy and time for
classification for each of these datasets is similar, and only a
representative information is shown in Table~\ref{tab:testacc}. It is
also worthwhile noting that the implementation of the original
algorithm is near optimal, as it scales well with increasing number of
processes. 

The results for USPS dataset, as shown in Figure~\ref{fig:usps} show the
efficacy of highly aggressive Multi2 heuristic, with Multi5pc being the
second best. These results validate our premise, that
$\left|\zeta\right|$ is
typically small, and a multiple 5\% heuristic such as Multi5pc can
provide significant elimination of computation for SVM, resulting in
faster convergence. As discussed previously, the first $\gamma$
reconstruction is executed at $20 \cdot \epsilon$, while others are
executed $2 \cdot \epsilon$. However,
with Multi* heuristics, the number of times the gradient is
reconstructed 
at the terminating condition can be predicted {\em apriori}. As shown in the USPS results, each of the Multi*
shrinking heuristics, although spend significantly more time in
gradient reconstruction(~\ref{algo:fcachereconst}), still reduce the overall execution time. 
For USPS
dataset, an overall speedup of $\approx$ 1.7x is observed in comparison to
Original implementation, and 5x in comparison to \texttt{LIBSVM}.

Figure~\ref{fig:mnist} shows the performance of various approaches on
MNIST dataset using 256 and 512 processes - equivalent of 16 and 32
compute nodes. There are several take away messages - the original
implementation scales well providing about 1.2x speedup or $\approx$
90\% efficiency. Several Multi* heuristics perform very well, with
little difference in execution time among them. For 256 processes, a
speedup of more than 1.3x is achieved using Multi1000 approach in
comparison to the original approach and 26x over \texttt{LIBSVM}. The fact that more time is spent in
$\gamma$ reconstruction is outweighed by the overall reduction in the
training time. 
Similar trends are observed with the w7a dataset
shown in Figure~\ref{fig:w7a}, where 2.3x speedup is observed on 32
processes and 1.6x speedup is observed on 64 processes, with up to 3.3x
speedup in comparison to \texttt{LIBSVM}.  

Figure~\ref{fig:mushrooms} shows the performance on the Mushrooms dataset.
In comparison to other datasets Mushrooms dataset requires
significantly more relative time for training due to the higher values
of $C$ and $\sigma^2$. As a result, the reconstruction time is relatively small.
$\frac{276}{8124}$ are support vectors,
which is less than 5\% of the overall training set.
Here as well,
Multi5pc provides near-optimal performance resulting in $\approx$ 3x
speedup, while Multi2 is slightly better than that. Again, it is fair to
conclude that Multi5pc is a good heuristic in extracting the benefits of
shrinking. Up to 8.2x improvement is observed in comparison to
\texttt{LIBSVM}.

Figure~\ref{fig:ijcnn1} shows the performance of IJCNN data set. We
have used this as an example to indicate that shrinking is not
beneficial for all datasets and different setting of hyperparameters.
For several datasets, we have observed that higher values of
hyperparameters results in faster elimination of samples, potentially
providing benefits of shrinking. This opens up a new avenue for research
where shrinking is integrated in the cross-validation step to get parameters
suitable for both shrinking and better generalization. As shown in the figure, the original
implementation is the best, while each of the other approaches result in
significant degradation due to shrinking. However, in comparison to
\texttt{LIBSVM}, a speedup of up to 9x is observed with the Original
implementation.


\section{Related Work}
\label{sec:related}
We discuss SVM training algorithms in literature under two major branches of study: 
$1)$ the sample selection methods and $2)$ parallel algorithms.

\subsection{Sample Selection}
Multiple researchers have proposed algorithms for selection of samples,
which can be be used for faster convergence. 
Active set methods solve the dual optimization problem by considering a
part of the dataset in a given iteration until global
convergence~\cite{plattsmo, joachims, keerthi, cao:2006:psm, fsv, psv,
cotter:gat}. The primary approach is to decompose large Quadratic Programming tasks
into small ones. Other approaches include the reformulation of the
optimization problem, which does not require the decomposition~\cite{olvi:rsvm}. 
The seminal SMO~\cite{plattsmo} and SVM$^{light}$~\cite{joachims} are active set
sequential methods and SVM-GPU~\cite{fsv}, and PSMO~\cite{cao:2006:psm}
are examples of parallel decomposition methods whereas
Woodsend {\em et al.}\cite{woodsend:2009:hmp} is an example of a parallel non-decomposition
solution. A primary problem with the working set methods is the
inability to address noisy, non-separable
datasets~\cite{yomtovparallel}. However, the simplicity, ease of
implementation and strong convergence properties make them an attractive choice for
solving large-scale classification problems.
 Other researchers have considered different values ($>2$) of
the working set~\cite{cotter:gat, joachims}.


\subsection{Parallel Algorithms}
\label{sec:appr}
With the advent of multi-core systems and cluster computing, several
parallel and distributed algorithms have been proposed in literature.
This section provides a brief overview of these algorithms.

Architecture specific solutions such as GPUs~\cite{fsv, cotter:gat} have been
proposed, and other approaches require a special cluster
setup~\cite{psv}.
Graf {\em et al.} have proposed Cascade SVM~\cite{psv}, which provides a
parallel solution to the dual optimization problem. The primary approach
is to divide the original problem in completely independent sub-problems,
and recursively combine the independent solutions to obtain the final
set of support vectors.  
However, this approach suffers from load
imbalance, since many processes may finish their individual
sub-problem before others. As a result, this approach does not scale well
for very large scale processes - a primary target of our approach.


The advent of SIMD architectures such as GPUs has resulted in research
conducted for Support Vector Machines on GPUs~\cite{fsv}. Under this
approach, a thread is created for each data point in the training set and
the MapReduce paradigm is used for compute-intensive steps. The primary
approach proposed in this paper is suitable for large scale systems, and
not restricted to GPUs. 

Several researchers have proposed alternative mechanisms for solving QP
problems. An example of variable projection method is proposed by 
Zanghirati and Zanni \cite{zanghirati03}. They use an iterative solver for QP problems leveraging
the decomposition strategy of SVM$^{light}$~\cite{joachims}. 
Chang {\em et al.} ~\cite{chang_psvm} have also considered more than 2 active set
size and solves the problem using Incomplete Cholesky Factorization and Interior
Point method (IPM).
Woodsend {\em et al.}~\cite{woodsend:2009:hmp} have proposed parallelization of linear SVM
using IPM and a combination of MPI and OpenMP. However, their approach is not an
active set method, as it does not decompose a large problem into
smaller ones. There are approaches like~\cite{pegasos} that solve the primal problem
for linear SVMs for very large problems, but the primary objective of
this paper is to scale the
most popular 2-working set based methods due to their ubiquity. 

As evident from the literature study above, none of the previously
proposed approaches use adaptive elimination of samples on large scale
systems, which has a significant potential in reducing the execution
time for several datasets.



\section{Conclusions and Future Work}
\label{sec:conclusions}
This paper has endeavored to address the limitations of previously proposed
approaches and provided a novel parallel Support Vector Machine
algorithm with adaptive
shrinking. It explored various design aspects of the algorithm and the associated
implementation, such as space complexity reduction by using sparse data
structures, intuitive heuristics for adaptive shrinking of
samples, and adaptive reconstitution of the data structures. 
We have used state-of-the-art
programming models such as Message Passing Interface (MPI)~\cite{mpi1}
and Global Arrays~\cite{vishnu:hipc12} for the design of communication and
data storage in the implementations. Empirical evaluation has demonstrated the efficacy of
our proposed algorithm and the heuristics.

The future work involves shrinking with second order heuristics for working
set selection, with
deeper evaluation of heuristics and, considering other algorithms and
working set sizes for faster elimination of samples. It will also be
interesting to study shrinking under different architectures like GPUs. Though the proposed
approach does complete elimination of kernel cache, it
is possible to use deep memory hierarchy for keeping active portions of
the kernel cache. The future work would also involve optimizations on
upcoming architectures such as Intel MIC architecture, and AMD Fusion
APU architecture.

\bibliographystyle{abbrv}
\bibliography{psmo,vishnu,vishnu_ml}

\begin{thebibliography}{10}

\bibitem{ml:climate}
P.~Balaprakash, Y.~Alexeev, S.~A. Mickelson, S.~Leyffer, R.~L. Jacob, and A.~P.
  Craig.
\newblock Machine learning based load-balancing for the cesm climate modeling
  package.
\newblock 2013.

\bibitem{exascale:report08}
K.~Bergman, S.~Borkar, D.~Campbell, W.~Carlson, W.~Dally, M.~Denneau,
  P.~Franzon, W.~Harrod, J.~Hiller, S.~Karp, S.~Keckler, D.~Klein, R.~Lucas,
  M.~Richards, A.~Scarpelli, S.~Scott, A.~Snavely, T.~Sterling, R.~S. Williams,
  K.~Yelick, K.~Bergman, S.~Borkar, D.~Campbell, W.~Carlson, W.~Dally,
  M.~Denneau, P.~Franzon, W.~Harrod, J.~Hiller, S.~Keckler, D.~Klein, P.~Kogge,
  R.~S. Williams, and K.~Yelick.
\newblock Exascale computing study: Technology challenges in achieving exascale
  systems peter kogge, editor and study lead, 2008.

\bibitem{burges}
C.~J.~C. Burges.
\newblock A tutorial on support vector machines for pattern recognition.
\newblock {\em Data Min. Knowl. Discov.}, 2:121--167, June 1998.

\bibitem{cao:2006:psm}
L.~J. Cao, S.~S. Keerthi, C.-J. Ong, J.~Q. Zhang, U.~Periyathamby, X.~J. Fu,
  and H.~P. Lee.
\newblock Parallel sequential minimal optimization for the training of support
  vector machines.
\newblock {\em IEEE Transactions on Neural Networks}, 17(4):1039--1049, July
  2006.

\bibitem{fsv}
B.~Catanzaro, N.~Sundaram, and K.~Keutzer.
\newblock Fast support vector machine training and classification on graphics
  processors.
\newblock In {\em Proceedings of the 25th international conference on Machine
  Learning}, ICML '08, pages 104--111. ACM, 2008.

\bibitem{libsvm}
C.-C. Chang and C.-J. Lin.
\newblock {LIBSVM}: A library for support vector machines.
\newblock {\em ACM Transactions on Intelligent Systems and Technology},
  2:27:1--27:27, 2011.
\newblock Software available at \url{http://www.csie.ntu.edu.tw/~cjlin/libsvm}.

\bibitem{chang_psvm}
E.~Y. Chang, K.~Zhu, H.~Wang, H.~Bai, J.~Li, Z.~Qiu, and H.~Cui.
\newblock Psvm: Parallelizing support vector machines on distributed computers.
\newblock In {\em NIPS}, 2007.
\newblock Software available at \url{http://code.google.com/p/psvm}.

\bibitem{cotter:gat}
A.~Cotter, N.~Srebro, and J.~Keshet.
\newblock A {GPU}-tailored approach for training kernelized {SVM}s.
\newblock In {\em Proceedings of the 17th ACM SIGKDD international conference
  on knowledge discovery and data mining}, KDD '11, pages 805--813, 2011.

\bibitem{nello}
N.~Cristianini and J.~Shawe-Taylor.
\newblock {\em An introduction to support vector machines: and other
  kernel-based learning methods}.
\newblock Cambridge University Press, 2000.

\bibitem{data:ascac13}
{DOE ASCAC Subcommittee}.
\newblock {Synergistic Challenges in Data-Intensive Science and Exascale
  Computing}, 2013.

\bibitem{dongarra}
J.~Dongarra.
\newblock Sparse matrix storage formats.
\newblock In Z.~Bai, J.~Demmel, J.~Dongarra, A.~Ruhe, and H.~van~der Vorst,
  editors, {\em Templates for the Solution of Algebraic Eigenvalue Problems: A
  Practical Guide}. SIAM, Philadelphia, 2000.

\bibitem{fletcher}
R.~Fletcher.
\newblock {\em Practical methods of optimization; (2nd ed.)}.
\newblock Wiley-Interscience, New York, NY, USA, 1987.

\bibitem{mpi2}
A.~Geist, W.~Gropp, S.~Huss-Lederman, A.~Lumsdaine, E.~L. Lusk, W.~Saphir,
  T.~Skjellum, and M.~Snir.
\newblock {MPI}-2: Extending the message-passing interface.
\newblock In {\em Euro-Par, Vol. I}, pages 128--135, 1996.

\bibitem{psv}
H.~P. Graf, E.~Cosatto, L.~Bottou, I.~Durdanovic, and V.~Vapnik.
\newblock Parallel support vector machines: The cascade svm.
\newblock In {\em Advances in Neural Information Processing Systems}, pages
  521--528. MIT Press, 2005.

\bibitem{mpi1}
W.~Gropp, E.~Lusk, N.~Doss, and A.~Skjellum.
\newblock {A High-Performance, Portable Implementation of the {MPI} Message
  Passing Interface Standard}.
\newblock {\em Parallel Computing}, 22(6):789--828, 1996.

\bibitem{hsieh}
C.-J. Hsieh, K.-W. Chang, C.-J. Lin, S.~S. Keerthi, and S.~Sundararajan.
\newblock A dual coordinate descent method for large-scale linear svm.
\newblock In {\em Proceedings of the 25th International Conference on Machine
  Learning}, ICML '08, pages 408--415, New York, NY, USA, 2008. ACM.

\bibitem{joachims}
T.~Joachims.
\newblock Making large-scale support vector machine learning practical.
\newblock In B.~Sch\"{o}lkopf, C.~J.~C. Burges, and A.~J. Smola, editors, {\em
  Advances in kernel methods}, pages 169--184. MIT Press, 1999.

\bibitem{keerthi}
S.~S. Keerthi, S.~K. Shevade, C.~Bhattacharyya, and K.R.K.Murthy.
\newblock Improvements to platt's smo algorithm for svm classifier design.
\newblock {\em Neural Computation}, 13(3):637--649, 2001.

\bibitem{nieplocha:psw94}
J.~Nieplocha, R.~J. Harrison, and R.~J. Littlefield.
\newblock {Global Arrays: A Nonuniform Memory Access Programming Model for
  High-Performance Computers}.
\newblock {\em Journal of Supercomputing}, 10(2):169--189, 1996.

\bibitem{plattsmo}
J.~C. Platt.
\newblock Fast training of support vector machines using sequential minimal
  optimization.
\newblock In {\em Advances in kernel methods: support vector learning}, pages
  185--208, Cambridge, MA, USA, 1999. MIT Press.

\bibitem{data:ascac11}
{Report from the DOE ASCR 2011 Workshop on Exascale Data Management, Analysis,
  and Visualization}.
\newblock {Scientific Discovery at the Exascale}, 2011.

\bibitem{pegasos}
S.~Shalev-Shwartz, Y.~Singer, and N.~Srebro.
\newblock Pegasos: Primal estimated sub-gradient solver for svm.
\newblock In {\em Proceedings of the 24th International Conference on Machine
  Learning}, ICML '07, pages 807--814, New York, NY, USA, 2007. ACM.

\bibitem{STOMP:homepage}
{Subsurface Transport over Multiple Phases}.
\newblock {STOMP}.
\newblock http://stomp.pnl.gov/.

\bibitem{tarca:bio07}
A.~L. Tarca, V.~J. Carey, X.-w. Chen, R.~Romero, and S.~Drăghici.
\newblock Machine learning and its applications to biology.
\newblock {\em PLoS Comput Biol}, 3(6):e116, 06 2007.

\bibitem{nwchem}
M.~Valiev, E.~Bylaska, N.~Govind, K.~Kowalski, T.~Straatsma, H.~V. Dam,
  D.~Wang, J.~Nieplocha, E.~Apra, T.~Windus, and W.~de~Jong.
\newblock Nwchem: A comprehensive and scalable open-source solution for large
  scale molecular simulations.
\newblock {\em Computer Physics Communications}, 181(9):1477 -- 1489, 2010.

\bibitem{vishnu:hipc12}
A.~Vishnu, J.~Daily, and B.~Palmer.
\newblock {Scalable PGAS Communication Subsystem on Cray Gemini Interconnect}.
\newblock Pune, India, 2012. HiPC.

\bibitem{vishnu:cass13}
A.~Vishnu, D.~J. Kerbyson, K.~Barker, and H.~J. J.~V. Dam.
\newblock Designing scalable pgas communication subsystems on blue gene/q.
\newblock Boston, 2013. 3rd Workshop on Communication Architecture for Scalable
  Systems.

\bibitem{vossen:hep08}
A.~Vossen.
\newblock {Support vector machines in high-energy physics}.
\newblock 2008.

\bibitem{woodsend:2009:hmp}
K.~Woodsend and J.~Gondzio.
\newblock Hybrid mpi/openmp parallel linear support vector machine training.
\newblock {\em J. Mach. Learn. Res.}, 10:1937--1953, Dec. 2009.

\bibitem{yomtovparallel}
E.~Yom-tov.
\newblock A parallel training algorithm for large scale support vector
  machines.
\newblock {\em Neural Information Processing Systems Workshop on Large Scale
  Kernel Machines}, 2004.

\bibitem{olvi:rsvm}
L.~Yuh-jye and O.~L. Mangasarian.
\newblock {RSVM}: Reduced support vector machines.
\newblock Technical Report 00--07, Data Mining Institute, Computer Sciences
  Department, University of Wisconsin, 2001.

\bibitem{zanghirati03}
G.~Zanghirati and L.~Zanni.
\newblock A parallel solver for large quadratic programs in training support
  vector machines.
\newblock {\em Parallel Computing}, 29(4):535--551, 2003.

\end{thebibliography}
\end{document}